\documentstyle [12pt]{article}
\title{
\hspace{3.0truein}{\small IFP-505-UNC}\\
\vspace{0.2truein}
{A Generalization of Quantum Statistics}
}
\author{Wei Chen\footnotemark[1]~,
 Y. Jack Ng\footnotemark[2]~, and Hendrik van Dam \\
Institute of Field Physics, Department of Physics
and Astronomy\\
University of North Carolina,
Chapel Hill, NC 27599-3255}
\date{
}
\footnotetext{$^*$ chen@physics.unc.edu}
\footnotetext{$^\dagger$  ng@physics.unc.edu}
\begin{document}

\maketitle
\vspace{0.2truein}
\begin{abstract}

	We propose a new fractional statistics for arbitrary dimensions,
based on an extension of Pauli's exclusion principle, to allow for
finite multi-occupancies of a single quantum state.  By explicitly
constructing the many-body Hilbert space, we obtain a new algebra of
operators and a new thermodynamics. The new statistics
is different from fractional exclusion statistics; and in a certain
limit, it reduces to the case of parafermi statistics.

\end{abstract}
\newpage
\baselineskip=22.0truept

	The thermodynamics of a macroscopic system is determined
microscopically by the statistics of its constituent particles and
elementary excitations.  Herein lies a fundamental significance of
statistics.  Ever since Heisenberg's second paper
on matrix mechanics, it has been known that a many-body wavefunction is
symmetric under permutations of identical bosons, but it is antisymmetric
for identical fermions.  The corresponding commutation and anticommutation
relations bilinear in field operators result in bose and fermi statistics
respectively.  Particles are accordingly classified into bosons and
fermions.  The overriding difference between the two groups is that
bosons condense while fermions exclude.  But it is natural to inquire
whether
there are any meaningful generalizations of statistics intermediate
between these two.

Attempts to generalize statistics dates back at least to Green's work
in 1953 \cite{Green}\cite{CK}. Green found that the principles of
quantum mechanics also allow
two kinds of statistics called parabose statistics and parafermi statistics of
positive integral order M (the M=1 cases reduce to the familiar Bose-Einstein
statistics and Fermi-Dirac statistics respectively).
They are described by trilinear commutation relations among the creation and
annihilation operators. Subsequently, the case of non-integral M was
investigated for possible deviations from Bose and Fermi statistics,
and in particular, for possible violations of Pauli's
exclusion principle \cite{IK}\cite{BTD}. This saga culminated with
a recent study of infinite statistics \cite{GM} in which all
representations of the
symmetric group can occur; this case is realized by the q-mutator algebras.

Another type of interpolating statistics, spearheaded by Wilczek,
is provided by the concept of anyons \cite{anyon}.
Anyons are particles whose
wavefunctions acquire an arbitrary phase when two of them are braided;
they obey fractional statistics.  More recently, Haldane
introduced another definition of statistics based on a generalization
of the Pauli principle  \cite{Hal}\cite{Wu}.
Unlike the anyon fractional exchange statistics
which is meaningful only in two spatial dimensions,
Haldane's fractional exclusion statistics is formulated in arbitrary
dimensions. The thermodynamics based on exclusion statistics is studied in
Ref.\cite{Wu}. The issue whether
anyons obey fractional exclusion statistics in the framework of quantum
field theory is addressed in Ref.\cite{CN}.  In this Letter we introduce
another type of statistics which is based on a (different) logical
extension of the
Pauli principle.  In the new statistics, we allow for multi-occupancy of a
single quantum state by up to a maximum number M of identical particles.
The M $=1$ case yields the conventional Fermi statistics while
the M $\rightarrow \infty$ case corresponds to the conventional
Bose statistics.  The new statistics bears some resemblance to fractional
exclusion statistics but is distinct from it.

In the literature a frequently used approach to study the quantum features of
(especially fermi and fermi-like) many-body systems is to start with
assumptions of commutation relations among certain operators. In this letter,
we would like to start with the construction of Hilbert space of quantum
states and then ``derive" some relations of the operators.
Consider the Hilbert space spanned by the eigenstates of the particle
number operator $\hat{N}$, \{$|j>, ~j = 0,1,2, \cdot\cdot\cdot\}$.
The $j$-particle state $|j>$ satisfies
\begin{equation}
\hat{N}|j> = j|j>\;, ~~~~
<j|k> = \delta_{jk}\;. \label{oth}
\end{equation}
Obviously, the ground state is the zero-particle state $|0>$.
We define the one-particle state as a %mixture
superposition of M single states $|e_i>$:
\begin{equation}
|1>=c_1c_2\cdot\cdot\cdot c_{{\rm M}-1}|e_1>
+ s_1c_2\cdot\cdot\cdot c_{{\rm M}-1}|e_2>+\cdot\cdot\cdot
+s_{{\rm M}-1}|e_{\rm M}>\;,\label{onep}
\end{equation}
where $c_i = {\rm cos}\theta_i$, $s_i={\rm sin}\theta_i$,
$\theta_i\neq0$, or multiples of $\pi/2$.
The use of the angles is just a matter of convenience. In this case,
the one-particle state can be thought of as  a unit vector on a
(M$-1$)-sphere. The  M single states $|e_i>$,
$i = 1,2,\cdot\cdot\cdot$ M, satisfy
\begin{equation}
<e_i|e_j> = \delta_{ij}\;,~~{\rm and}~~ (|e_i>)^2=0\;.\label{pauli}
\end{equation}
The latter condition can be understood as
a reflection of the Pauli exclusion principle, namely, a two-$|e_i>$
state is forbidden.

The two-particle state is defined as a superposition
of the tensor products \newline $|e_i>|e_j>$ %($i\neq j$)
\begin{equation}
|2> = \sum^{\rm M}_{i<j}c_{ij}|e_i>|e_j>\;.
\end{equation}
Similarly, three and more particle states can be defined
as a superposition of tensor products  $|e_i>|e_j>|e_k>$ with
c-number coefficients $c_{ijk}$ and so on.
How to determine $c_{ij}$, $c_{ijk}$ and so on,
in term of the angles $\theta_i$, %amplitudes $f_j$'s,
will be discussed later. We should
emphasize that in a tensor product of single states, we adopt the rule
that the order of single states does not make a difference, {\it i.e.}
$|e_i>|e_j>=|e_j>|e_i>$. Due to the second condition in Eq.(\ref{pauli}),
the M-particle state
\begin{equation}
|{\rm M}>=\prod_{i=1}^{\rm M}|e_i>\;,
 \end{equation}
is the maximal-particle state, the state with the maximum number of
M particles. There are no states beyond it, {\it i.e.} states
$|j>$ for $j>$ M do not exist.

Now, we introduce the annihilation
and creation operators, $a$ and $a^\dagger$,
\begin{equation}
a|j>=f_j|j-1>\;,~~~~
a^\dagger|j>=f^*_{j+1}|j+1>\;.\label{creat}
\end{equation}
In particular, $a|0>=0$ and $a^\dagger|0>=|1>$.
Hereafter we choose $f_1$ and all $f_j$ to be real for convenience
(this does not affect
the physics we will discuss in the single particle species case).
One can readily check the following
commutation relation of $\hat{N}$ with $a$ and $a^\dagger$
\begin{equation}
\hat{N} a - a \hat{N} = - a\;,
{}~~{\rm and}~~ \hat{N} a^\dagger - a^\dagger \hat{N} = a^\dagger\;.
\end{equation}

Starting from the one-particle state Eq.(\ref{onep}), using the relation
\begin{equation}
(a^\dagger)^j|0> = (|1>)^j\sim |j>\;, ~~{\rm or}~~
 a^j|j> \sim |0>\;,
\end{equation}
the normalization condition in Eq.(\ref{oth}),
and $f_0 = 0$ and $f_1 = 1$, one can systematically determine the
amplitude $f_2$ and the coefficients $c_{ij}$, $f_3$ and $c_{ijk}$,
and so on, set by set.
To be concrete, let us take ${\rm M} = 2$ and $3$ as examples.
For ${\rm M}=2$, the one-particle state is $|1> = c|e_1>+s|e_2>$.
Acting the creation operator $a^\dagger$ on $|1>$, one has
$a^\dagger|1>=(|1>)^2=2cs|e_1>|e_2>=f_2|2>$.
Therefore $f_2=2{\rm cos}\theta{\rm sin}\theta$.
Obviously $|2>$ is the maximal-particle state since
$|3>=0$. This is tantamount to $f_j =0$ for $j > 2$, or equivalently
$(a^\dagger)^3 = a^3 = 0$.

For ${\rm M}=3$, the one-particle state is
$|1> = c_1c_2|e_1>+s_1c_2|e_2>+s_2|e_3>$. From
$a^\dagger|1>=(|1>)^2=f_2|2>$,
we obtain
$f_2 = 2s_1c_1c_2^2\sqrt{1+t_2^2/s^2_1c_1^2}$
with $t_2 = {\rm tan}\theta_2$ and
$|2>= 1/\sqrt{1+t_2^2/s_1^2c_1^2}
 \left(|e_1>|e_2>+t_2/s_1|e_1>|e_3>
+t_2/c_1|e_2>|e_3>\right)$.
Using $a^\dagger|2>=f_3|3>$ with
$|3>=|e_1>|e_2>|e_3>$, we readily obtain
$f_3=3s_2/\sqrt{1+t_2^2/s^2_1c^2_1}$.
The amplitudes $f_j = 0$ for $j>3$,
equivalently $(a^\dagger)^4 = a^4 = 0$.

Let us consider the operator algebra of $a$ and $a^\dagger$.
First, for a general M, there are no simple bi-linear
operator relations like $[a,a^\dagger]_\pm = 1$ that
the conventional boson and fermion operators satisfy.
Instead, here we have
\begin{equation}
(aa^\dagger + a^\dagger a)|j> = (f^2_j+f^2_{j+1})|j>\;.
\end{equation}
For ${\rm M}=1$, we recover the anti-commutation
relation for fermions.

For any given M, $a^j = (a^\dagger)^j = 0$ only if $j>$ M.
Therefore, there exist (M+1)-linear relations between $a$
and $a^\dagger$. For example, for M $=2$, one has the cubic
relations
\begin{eqnarray}
a^2a^\dagger+f_2^2a^\dagger a^2 &=& f_2^2a\;,\label{cub1}\\
a^2a^\dagger+a^\dagger a^2+aa^\dagger a
&=&{\rm Tr}(aa^\dagger)a\;,\label{cub2}
\end{eqnarray}
where ${\rm Tr}(aa^\dagger) = \sum_{j=1}^{\rm M} f_j^2$,
plus the hermitian conjugate relations.

For ${\rm M}=3$, there are the quartic relations like
\begin{eqnarray}
a^3a^\dagger + f_3^2a^\dagger a^3
&=& f_3^2a^2\;,\\
a^3a^\dagger+a^\dagger a^3+a^2a^\dagger a+aa^\dagger a^2
 &=&{\rm Tr}(aa^\dagger)a^2\;,
\end{eqnarray}
plus the hermitian conjugate relations.
We speculate that the multi-linear relations among $a$ and $a^\dagger$ in
the limit M $\rightarrow \infty$ (and with a suitable choice of the
theta angles) actually reduce to the bilinear commutation
relation for Bose statistics.

Next we consider the particle number operator $\hat{N}$ in terms
of the creation and annihilation operators $a^\dagger$ and  $a$.
One way to do this is to ssume, for a given M,
\begin{equation}
\hat{N} = C_1a^\dagger a + C_2 (a^\dagger)^2a^2
+ \cdot \cdot \cdot + C_{\rm M} (a^\dagger)^{\rm M}a^{\rm M}\;.
\label{N}
\end{equation}
Then the M coefficients $C_j$, $j= 1,2,\cdot\cdot\cdot, {\rm M}$,
can be determined
by using the M independent equations
$\hat{N}|k>=k|k>$, $k=1,2, \cdot\cdot\cdot, {\rm M}$.
For example, for M $=2$: $C_1=1$ and $C_2=(2-f_2^2)/f_2^2$.
For M $=3$:  $C_1=1$, $C_2=(2-f_2^2)/f_2^2$, and
$C_3=(3/f_3^2-3+f_2^2)/f_2^2$.

The particle number operator $\hat{N}$ can be expressed in
other forms for certain values of the theta angles.
For example, if the $\theta_i$'s in Eq.(\ref{onep})
are chosen so that the one particle state is
$|1> = 1/\sqrt{\rm M}\left(|e_1>+|e_2>+\cdot\cdot\cdot|e_{\rm M}>\right)$
and if the operators $a$ and $a^\dagger$ are replaced by operators
$b/\sqrt{\rm M}$ and $b^\dagger/\sqrt{\rm M}$, the particle number operator
takes the form  $\hat{N} = \frac{1}{2}(b^\dagger b - bb^\dagger)
+\frac{{\rm M}}{2}{\bf 1}$. This latter form was used in the study of
parafermi statistics  \cite{Green}\cite{CK}.

With the %construction of the
Hilbert space of quantum states for the new statistics
and the diagonal particle number operator now available, it is natural and
straightforward to consider the
quantum statistical mechanics of a system that is compatible
with such a construction (note that systems of this kind
 are not necessarily described by a free theory).

Let us assume a single particle to have energy
$\epsilon = \epsilon({\bf p})$, then the Hamiltonian operator takes the form
\begin{equation}
\hat{H} = \epsilon\hat{N}\;.\label{H}
\end{equation}
The energy spectrum of the system for a given M is given by
\{$\epsilon, ~2\epsilon,~\cdot\cdot\cdot,~{\rm M}\epsilon$\},
similar to that for a spin system in a magnetic field.

The grand partition function is %for a given M is
\begin{equation}
{Z} = {\rm Tr} e^{-\beta(\hat{H}-\mu\hat{N})}
 = \sum_{j=0}^{\rm M} <j| e^{-\beta(\hat{H}-\mu\hat{N})}|j>
= \prod_{\bf p} \sum_{j=0}^{\rm M}(ze^{-\beta\epsilon})^j
\;,
\end{equation}
where $\beta$ is the reciprocal of temperature $T$, $\mu$ is the
chemical potential, the fugacity is $z=e^{\beta\mu}$,
and the Boltzmann constant is $k_B=1$. For ${\rm M}=1$,
${Z} = \prod_{\bf p}(1+ ze^{-\beta\epsilon})$
is the partition function for free fermions; while
for ${\rm M}=\infty$, ${Z} = \prod_{\bf p}1/(1- ze^{-\beta\epsilon})$
recovers the partition function of free bosons \cite{LL}.
For M $\neq 1$ and $\infty$, $Z$ describes a system interpolating
between free fermions and free bosons.
Note that the resulting thermodynamics is insensitive to
the particular values of the set of amplitudes $f_j$. In particular,
for the M $=2$ case, even when the one-particle state
$|1>$ is predominantly $|e_1>$ or $|e_2>$, the statistics is
very different from Fermi statistics, suggesting that the
models discussed in Ref.\cite{IK} do not yield weak violations
of the Pauli principle as correctly pointed out in %Biedenharn et al.
Ref.\cite{BTD}.

With the grand partition function $Z$,
one can calculate various thermodynamical
quantities. The particle number %distribution function is given by
\begin{equation}
N = z\frac{\partial}{\partial z} {\rm ln}Z
= \sum_{\bf p}\frac{\sum_{j=1}^{\rm M}j(ze^{-\beta\epsilon})^j}
{\sum_{j=0}^{\rm M}(ze^{-\beta\epsilon})^j}\;.
\label{NN}
\end{equation}
Accordingly, the average occupation numbers are %$n(\epsilon)$ are given by
\begin{equation}
n(\epsilon) = \frac{\sum_{j=1}^{\rm M}j(ze^{-\beta\epsilon})^j}
{\sum_{j=0}^{\rm M}(ze^{-\beta\epsilon})^j}\;.\label{distri}
\end{equation}
At $T = 0$, $n(\epsilon) = 0$ for $\epsilon > \mu$;
while  $n(\epsilon) = {\rm M} $ for $\epsilon < \mu$. The fermi energy
$\epsilon_F$ is defined by the particle density
$n=N/V=(1/V)\sum_{{\bf p}<{\bf p}_F}n(\epsilon)$ at absolute zero.
As $T \rightarrow \infty$, $n(\epsilon)={\rm M}/2$.

One can also calculate the entropy $S$ by applying
$S=-\partial F/\partial T,$
where $F=-T{\rm ln}Z$ is the grand potential. Using Eq.(\ref{distri}) to
invert $ze^{-\beta\epsilon}$ in term of the average occupation number
$n(\epsilon)$ one can then express $S$ in term of $n(\epsilon)$.
For example, for M $=2$, we find
\begin{equation}
S = \sum_{\bf p}\left(-{\rm ln}n(\epsilon)
+ (1-n(\epsilon)){\rm ln}x+{\rm ln}(1+2x)\right)\;,
\end{equation}
where $x=(\sqrt{1+6n(\epsilon)-3n^2(\epsilon)}-1
+n(\epsilon))/2(2-n(\epsilon))$.

The equation of state is given by %for a given M is
\begin{equation}
\beta PV = {\rm ln}Z = \sum_{\bf p}
{\rm ln}\sum_{j=0}^{\rm M}(ze^{-\beta\epsilon})^j
%\frac{1-z^{({\rm M}+1)} e^{-({\rm M}+1)\beta\epsilon}}
%{1-ze^{-\beta\epsilon}}
\;,\label{P}
\end{equation}
where $P$ denotes the pressure and $V$ the volume. In the large
volume ($V\rightarrow \infty$) limit, we replace the sum over momentum
${\bf p}$ by the integral over  ${\bf p}$: $\sum_{\bf p}$ $\rightarrow$
$V\int d^D{\bf p}/(2\pi)^D$. Such a replacement is clearly valid only if
the summand is finite for all ${\bf p}$.
For the bose gas (the limit of M $\rightarrow\infty$), the summand
 $-{\rm ln}(1-ze^{-\beta\epsilon})$ in Eq.(\ref{P})
diverges as the fugacity $z\rightarrow 1$, because
the single term corresponding to ${\bf p}=0$ diverges.
This is of course related to %in fact in connection with
the Bose-Einstein condensation. On the other hand, for
any finite M, the summand in Eq.(\ref{P}) is finite for
any value of $\epsilon({\bf p})$.

In all our discussions so far we have made no reference %not been restricted
to any specific spatial
dimensions. We now consider a planar system. Furthermore, we assume the
single particle energy $\epsilon$ take the form $\epsilon({\bf p})
={\bf p}^2/(2m)$, with $m$ being the (effective) mass of the particles or
excitations so that the system is an ideal gas.
Using Eq.(\ref{P}) and Eq.(\ref{NN}), and performing the integrations
over ${\bf p}$, we readily obtain
\begin{eqnarray}
\beta P&=&\frac{1}{\lambda^2}
\sum_{k=1}^\infty\frac{z^k}{k^2}\left(1-\frac{z^{{\rm M}k}}
{{\rm M}+1}\right)\; ~~~~~~~~~~~~~~~~~~~~~~~~~~~~~~~~~~~~~~~~~~
(z\leq 1),\label{PP}\\
&=&\frac{1}{\lambda^2}\left[\frac{\pi^2}{3}\frac{{\rm M}}{{\rm M}+1}
+\frac{{\rm M}}{2}({\rm ln}z)^2-
\sum_{k=1}^\infty\frac{z^{-k}}{k^2}\left(1-\frac{z^{-{\rm M}k}}
{{\rm M}+1}\right)\right]\; ~~~~~~~(z\geq 1),\label{PP1}\\
n&=&\frac{N}{V}=\frac{1}{\lambda^2}
{\rm ln}\frac{1-z^{({\rm M}+1)}}{1-z}\;,\label{nn}
\end{eqnarray}
where $\lambda = \sqrt{2\pi\beta/m}$ is the thermal wavelength.
Solving for $z$ in Eq.(\ref{nn}) and substituting it into
Eq.(\ref{PP}), in the high temperature and low density limit,
{\it i.e.} $\lambda^2 n\ll 1$, we can conduct a virial expansion
in the form $\beta P = n\left(1+B_2\lambda^2 n+B_3(\lambda^2 n)^2
+\cdot\cdot\cdot\right)$.
For M $=2$, we find $B_2= -1/4$, $B_3=25/36$, $\cdot\cdot\cdot$;
and for M $=3$,  $B_2= -1/4$, $B_3=1/36$, $\cdot\cdot\cdot$.
Actually from Eq.(\ref{PP}) and Eq.(\ref{nn}),
it is not difficult to check that for any M $>1$, the
second virial coefficient is $-1/4$, the same as for the
ideal bose gas (M $=\infty$). It implies that (for sufficiently small
$\lambda^2 n$) the quantum effect
on the ideal gas, for all M except M $=1$, is equivalent to an
attractive ``interaction'' among excitations.
For ideal fermi gas (M $=1$) this effect is a repulsive one as $B_2=+1/4$.

In the low temperature and high density limit, {\it i.e.} $\lambda^2 n\gg 1$,
most particles are in the states with energy
$\epsilon<\epsilon_F=2\pi n/{\rm M}m$. Using $U=F+TS+\mu N$, we obtain
the internal energy
\begin{equation}
U = \frac{1}{2}N\epsilon_F\left(1
+\frac{2\pi^2}{3({\rm M}+1)}(\frac{T}{\epsilon_F})^2+\cdot\cdot\cdot\right)\;.
\end{equation}
The first term is the ground state energy, a result that can be verified by
using $\sum_{{\bf p}<{\bf p}_F}{\rm M}{\rm p}^2/(2m) = N\epsilon_F/2$.
{}From the above equation the specific heat at constant volume can be readily
found: $C_V/N\simeq\frac{2\pi^2}{3({\rm M}+1)}\frac{T}{\epsilon_F}$.
These suggest that in the low temperature and high density limit, a system
in which each quantum state has a maximum multi-occupancy
of M $< \infty$ is like the
fermion system (M$=1$).

A comparison of the quantum statistical mechanics
obtained here with that for exclusion statistics can now be made.
Both fractional statistics, based on generalizations of
Pauli's exclusion principle, are well defined in
arbitrary dimensions. At zero temperature, the $n(\epsilon)$ distributions
in the two thermodynamics are the same if the statistical parameter
$g$ in exclusion statistics is identified with $1/$M in the new statistics.
But there are profound differences between the two statistics.
%For the case of fractional exclusion statistics,
The second virial coefficient for a free planar exclusion statistical system,
such as the free anyon system, is given by $B_2 = 1/4 - g/2$
(where $g=0, 1$ for fermion and boson statistics respectively)
\cite{Wu}\cite{CN}.
Accordingly the statistical interaction is attractive, neutral,
or repulsive, depending on the value of $g$.
In particular, for semions with $g=1/2$, it is neutral (to this order);
and for others with fractional $g=1/$M (M $>2$), it is repulsive.
In contrast, we find an attractive interaction for all M $> 1$ in
the new statistics discussed above. Moreover, the particle distribution
$n(\epsilon)$ are in general very different in the
two statistics. For example, the distribution of semions %($g=1/2$)
from the exclusion statistical derivation takes the form
\cite{Wu}: $n(\epsilon) = 1/\sqrt{1/4+e^{2\beta\epsilon}/z^2}$. This
is different  (except at zero temperature) from
the one given by Eq.(\ref{distri}) for M $=2$,
$n(\epsilon)=(2+e^{\beta\epsilon}/z)/(1+ e^{\beta\epsilon}/z
+e^{2\beta\epsilon}/z^2)$.
Integrating out the distributions over $\epsilon$, we find the
resulting densities are different too.
The statistical weights $W$ for fractional exclusion statistics
are also quite different from those for the new statistics.
For instance, for $N$ identical semions occupying G states,
it reads $W= \left(\begin{array}{c}
G+\frac{1}{2}(N-1)\\N\end{array}\right)$ in exclusion statistics
\cite{Hal}\cite{Wu}, whereas for M $=2$ in the new statistics we obtain
\begin{equation}
W=\sum_0^{[\frac{N}{2}]} \left(\begin{array}{c}
N-k\\k\end{array}\right) \left(\begin{array}{c}
G\\N-k\end{array}\right)\;,%~~~~~~~~~~~~~~~({\rm M}=2)\;,
\end{equation}
where $[\frac{N}{2}]$ denotes $N/2$ and $(N-1)/2$ for $N=$ even and odd,
respectively. We conclude that the statistics associated with
multi-occupancy of a single quantum state and the resulting operator
algebra are different from the exclusion statistics defined in Ref.\cite{Hal}
(in contrast to a recent proposal \cite{KN}, in which the exclusion
statistical parameter $g$ is assumed to
be connected to the maximum occupancy number M by $g=1/$M and the
statistical distributions in exclusion statistics are
connected to the amplitudes $f_j$).

We thank K. Dy, E. Merzbacher, V.P. Nair, G.W. Semenoff,
and Y.-S. Wu for useful conversations.
This work was supported in part by
the  U.S. DOE grant No. DE-FG05-85ER-40219.

\vspace{0.5cm}

{\it Note added}: After this paper was completed, we noticed a recent
interesting work \cite{Polych}, in which an issue relevant to this paper
was also addressed.

\end{document}